\documentclass[letterpaper]{article}

\newcommand{\be}{\begin{equation*}}
\newcommand{\ee}{\end{equation*}}
\newcommand{\bee}{\begin{eqnarray*}}
\newcommand{\eee}{\end{eqnarray*}}
\usepackage{bbm}
\usepackage{bm}
\usepackage{amsmath,amsfonts,amssymb,amsthm}
\usepackage{url}
\usepackage{graphicx}

\begin{document}

\title{Differentiable Phylogenetics via Hyperbolic Embeddings with Dodonaphy}

\author{Matthew Macaulay and Mathieu Fourment\\
Australian Institute for Microbiology \& Infection,\\
University of Technology Sydney,\\
Ultimo, 2007, NSW, Australia\\
mathieu.fourment@uts.edu.au
}

\maketitle

\abstract{
\textbf{Motivation:}
Navigating the high dimensional space of discrete trees for phylogenetics presents a challenging problem for tree optimisation.
To address this, hyperbolic embeddings of trees offer a promising approach to encoding trees efficiently in continuous spaces.
However, they require a differentiable tree decoder to optimise the phylogenetic likelihood.
We present soft-NJ, a differentiable version of neighbour-joining that enables gradient-based optimisation over the space of trees.
\\
\textbf{Results:}
We illustrate the potential for differentiable optimisation over tree space for maximum likelihood inference.
We then perform variational Bayesian phylogenetics by optimising embedding distributions in hyperbolic space.
We compare the performance of this approximation technique on eight benchmark datasets to state-of-art methods.
However, geometric frustrations of the embedding locations produce local optima that pose a challenge for optimisation.
\\
\textbf{Availability:}
Dodonaphy is freely available on the web at \url{www.https://github.com/mattapow/dodonaphy}.
It includes an implementation of soft-NJ.\\
\textbf{Contact:}
\url{mathieu.fourment@uts.edu.au}
}

\section{Introduction}
Phylogenetics provides us with the evolutionary history of a set of taxa given their genetic sequences, which is usually a bifurcating tree.
However, fast optimisation relies on gradients, which are not well defined between discrete trees.
Thus most tree optimisation techniques consider manual changes to the tree topology before optimising the continuous parameters (branch lengths) of each tree considered~\cite{minh2020iqtree, stamatakis2014raxml}.
Knowing which of the super-exponential number of trees to manually try is a challenging task~\cite{guindon2010new, ki2022variational}.

Providing a differentiable way to move between tree topologies, would allow well-developed continuous optimisation techniques to work in the space of phylogenetic trees.
In this paper, we propose a novel technique to continuously move through the space of bifurcating trees with gradients.
Our approach hinges on two ideas a) an embedding of the genetic sequences into a continuous space and b) an algorithm we propose called $\text{soft-NJ}$, which passes gradients through the neighbour joining algorithm.
With these preliminaries, we can embed the tip nodes of a tree in the continuous embedding space and then optimise the locations of these nodes based on the neighbour joining tree that they decode from $\text{soft-NJ}$.

We use hyperbolic embeddings to represent trees in a continuous manner.
This is similar to embedding points in Euclidean space, where each tip node of the tree is positioned in the space with a certain location~\cite{layer2017phylogenetic}.
However, the metric between two points is modified to give a negative curvature between (as opposed to positive curvature for points on a sphere).
Hyperbolic data embeddings offer low dimensional, efficient, and precise ways to embed hierarchically clustered data~\cite{peng2022hyperbolic, chami2020trees, monath2019gradientbased, nickel2017poincare, chami2020lowdimensional} or tree-like data in phylogenetics~\cite{koptagel2022vaiphy, wilson2021learning, corso2021neural, nagano2019wrapped, macaulay2023fidelity, iuchi2021representation}.

Alternative continuous tree embedding methods are high dimensional, growing significantly with increasing taxa; BHV space grows double factorially~\cite{billera2001geometry}, flattenings of sequence alignments grow exponentially~\cite{allman2008phylogenetic}, sub-flattenings increase quadratically~\cite{sumner2017dimensional}, as with tropical space~\cite{speyer2003tropical}.
In these spaces, each point corresponds to a single tree, making them high dimensional.
Additionally, they have non-differentiable boundaries between trees, making them difficult to optimise in~\cite{dinh2017probabilistic}.
Whereas with hyperbolic embeddings, each taxon has an embedding location and together the set of taxa locations decode to a tree.
This keeps the embedding space low dimensional and the number of optimisation parameters linear in the number of taxa.

The goal of our approach is to optimise the embedding locations with gradient-based optimisation, which requires a differentiable loss function (i.e. the likelihood or unnormalized posterior probability).
This is easily achieved in other applications with carefully designed loss functions.
However, in phylogenetics, there are well accepted Markov models of evolution (such as GTR or JC69), which rely on having a tree structure to compute their likelihood.
To maximise the likelihood by changing the embedding locations, we developed $\text{soft-NJ}$ --- a differentiable version of the neighbour joining algorithm using automatic differentiation.
It allows gradients to pass from the embedding locations into a decoded tree and the likelihood function.

We implemented $\text{soft-NJ}$ in Dodonaphy, a software for likelihood-based phylogenetics using hyperbolic space.
We demonstrate this newfound ability for phylogenetic optimisation with two modes of gradient based inference: maximum likelihood (ML) and Bayesian variational inference (VI).

Variational inference is a Bayesian technique for approximating the posterior distribution with simple and tractable distributions, as reviewed in~\cite{blei2017variational}.
It indirectly finds the variational distribution that minimises the KL divergence between the unnormalised posterior and the variational distribution.
This avoids the need to compute the normalising constant in Bayes theorem or to resort to time consuming Markov chain Monte Carlo sampling, potentially offering significant computational speed ups.

Recently, phylogenetic variational inference has garnered increasing attention~\cite{zhang2019variational, zhang2020improved, ki2022variational, koptagel2022vaiphy} as a promising way to cope with high dimensionality inherent to Bayesian phylogenetics.
Concurrently, variational approximations have extended to general manifolds, such as hyperbolic space, where the variational density sits on the manifold~\cite{wilson2018gradient, tran2021variational, peng2022hyperbolic}.
We combine these two paradigms to perform variational Bayesian phylogenetic inference on hyperbolic manifolds.

To perform variational inference on the space of phylogenies, we equip each of $n$ embedded taxon locations with a variational distribution (a projected multivariate-Normal) in hyperbolic space $\mathbb{H}^{d}$.
We optimise the set of $n$ probability distributions in hyperbolic space. 
We can quickly draw samples from these distributions and compute their neighbour joining tree of the sample.
This yields a distribution of phylogenetic trees that approximate the posterior distribution.

It's worth noting that $\text{soft-NJ}$ is not limited to phylogenetics; this contribution opens up a wide range of continuous gradient-based inference methods to any hierarchically structured data.
Recent advances in machine learning have also pushed for learning embeddings for hierarchical data such as in natural language processing~\cite{chami2020trees, monath2019gradientbased, nickel2017poincare}.
Recent machine learning problems also attempt to optimise tree structures.
\text{Soft-NJ} provides an alternative algorithm to search through the space of trees in a differentiable manner and need not be constrained to phylogenetic problems.
Additionally, a similar approach to \text{soft-NJ} could be applied to the UPGMA algorithm, which is used widely used outside of phylogenetics.

\section{System and methods}\label{sec:systems}
In this section, we provide the necessary background for our proposed phylogenetic embedding technique.
First, we recap how phylogenetic models are used for tree inference in maximum likelihood and Bayesian approaches, in particular, variational Bayesian inference.
We then introduce hyperbolic space and how phylogenies can be embedded in this space.

\subsection{Phylogenetic Inference}\label{sec:systems:phylogenetic}
Phylogenetic models compute the likelihood of an aligned set of genetic sequences $D$, which are observed at the tips given a bifurcating tree $T$~\cite{felsenstein1973maximum}.
Let $T = T(\tau, \ell_{\tau})$ denote an unrooted bifurcating tree with topology $\tau$ and continuous branch lengths $\ell_{\tau}$.
A phylogenetic model (denoted $\mathcal{M}$) is a Markov model between the four nucleotide states $A, C, G, T/U$ along the tree at each site in the alignment~\cite{tavare1986probabilistic}.
It has six substitution rates which sum to one and four equilibrium frequencies which also sum to one.
We use the GTR model and a simplified version of it called JC69~\cite{jukes1969evolution} to compute the likelihood of the alignment data $D$ given a tree $p(D | T, \mathcal{M})$.

\subsection{Bayesian Phylogenetic Models}\label{sec:systems:bayesian}
Bayesian phylogenetics includes prior knowledge of each parameter and seeks the posterior distribution over phylogenetic trees given a multiple sequence alignment.
The posterior is $p(T, \mathcal{M} |D) \propto p(D | T, \mathcal{M})p(T)p(\mathcal{M})$, with, in general, an unknown normalising constant.

We specify the prior probability of an unrooted tree $p(T)$ using a Gamma-Dirichlet model~\cite{rannala2012tail}.
The Gamma-Dirichlet prior invokes a Gamma distribution (shape $1$, rate $0.1$) over the total tree length before dividing this length into the branches with an equally weighted Dirichlet distribution~\cite{rannala2012tail}.
The GTR model's prior $p(\mathcal{M})$ is a flat Dirichlet for the six substitution rates and a flat Dirichlet on the four equilibrium frequencies.

\subsection{Variational Inference}
Variational inference minimises some measure of divergence between an approximating function $q$ from a family of distributions $q \in \mathcal{Q}$ and the posterior target $p(T, \mathcal{M} | D)$.
We use the standard KL-divergence between the two distributions, which after dropping the $\mathcal{M}$ and putting it in log space is:
\bee
\text{KL}\big(q(T)|| p(T|D\big) &=& \mathbb{E}[\log q(T)] - \mathbb{E}[\log p(T|D)]\\
&=& \mathbb{E}[\log q(T)] - \mathbb{E}[\log (p(D|T)) + \log(p(D))]
\eee
where the expectations are taken with respect to $q(T)$.
The marginal likelihood of the data $\log p(D)$ is intractable to compute, however, since the data is constant, we can simply drop this term and optimise to the same optimum.
As a result, the so called evidence lower bound (ELBO) becomes the objective to maximise:
\be
\mathcal{L}_{\text{ELBO}} = \mathbb{E}[\log p(T, D)] - \mathbb{E}[\log q(T)]
\ee
Maximising the \text{ELBO} is equivalent to minimising the KL-divergence between the target $p(T|D)$ and variational distributions $q(T)$ for any given data set.

\subsubsection{Improved VI}
The chosen variational distribution $q(T)$ may be too simple to capture the true posterior distribution, so to allow for more expressive variational distributions, they can be \textit{boosted} with a mixture model.
Boosting is the process of attaining stratified samples over multiple variational distributions $q_{k}(T)$ each with weight $\alpha_{k},\; k \in 1, 2, ... K$.
Each sample can be computed with $M$ importance samples as done in the stratified importance weighted auto-encoder (SIWAE)~\cite{morningstar2021automatic}:
\be
\mathcal{L}_{\text{SIWAE}} = \mathbb{E}_{q}\Big[\\
 \log \frac{1}{M} \sum_{m=1}^{M} \sum_{k=1}^{K} \alpha_{k} \dfrac{p(T, D)}{q_{k}(T)} \Big]
\ee
Compared to other objectives, this version of the ELBO has improved expressivity and encourages the mixtures not to collapse onto each other~\cite{morningstar2021automatic, burda2016importance}.
We optimise the parameters of the variational distribution to maximise the SIWAE.

Unless otherwise stated, we selected the hyper-parameters $M=1$ importance samples, $K=1$ boosts (mixtures) with equal initial weights $\alpha_{k}=1/K$.
We use PyTorch's Adam optimiser with a learning rate of $0.1$.
The learning rate decayed according to $(t + 1)^{-0.5}$, where $t$ is the iteration number.

\subsection{Hyperbolic Space}\label{sec:systems:hyperbolic}
We model $d$-dimensional hyperbolic space by a hyperboloid: 
\be
\mathbb{H}^{d} = \{u \in \mathbb{R}^{d+1}: \langle u, u\rangle = -1\},
\ee
where the Lorentz inner product is
\be
\langle u, v\rangle = -u_{0}v_{0}+u_{1}v_{1}+....+u_{d}v_{d}.
\ee
This is a sheet sitting in the ambient space $\mathbb{R}^{d+1}$.
The distance between two points on the sheet is
\be
d_{\kappa}(u, v) = \frac{1}{\sqrt{-\kappa}}\text{arcosh}(-\langle u, v\rangle),
\ee
where $\kappa<0$ is the curvature of the manifold.
Based on previous work, we select three dimensions $d=3$~\cite{macaulay2023fidelity}.

\subsection{Encoding Trees in $\mathbb{H}^{d}$}\label{sec:algorithm:encoding}
To initialise an embedding in hyperbolic space, we take a tip-tip distance matrix from a given phylogenetic tree: $D_{T}$.
Dodonaphy then uses Hydra+ to embed each taxon with a location $\vec{z}_{i}$ in hyperbolic space with $d$ dimensions $\vec{z}_{i}\in \mathbb{H}^{d}$.
Hydra+ is a recent adaption of multi-dimensional scaling to hyperbolic space~\cite{keller-ressel2020hydra}.
It is an optimisation algorithm that minimises the stress of the embedding, that is, it minimises the difference between the given distance matrix $D_{T}$ and the pairwise distances in hyperbolic space $D_{ij} = d_{\kappa}(\vec{z}_{i}, \vec{z}_{j})$.
The result is a set of embedding locations in $\vec{z}_{i} \in \mathbb{H}^{d}$, one for each tip $i$ in the phylogenetic tree.

Note that this is an approximate embedding technique, so an encoded tree may not decode back to the originally given tree.

\subsection{Encoding Tree Distributions in $\mathbb{H}^{d}$}
To encode a variational distribution over trees in hyperbolic space, each taxon requires a variational distribution in $\mathbb{H}^{d}$.
To initialise an embedding, for each taxon, we centred a distribution around the point $\vec{z}_{i}$ as in the previous section.
We set the covariance to be diagonal, i.e. mean-field, using a coefficient of variation of $20$ compared to the smallest tip-tip distance.

Each variational distribution is a multivariate Normal $\mathcal{N}(\mu, \Sigma)$ projected from the tangent space at $(1, 0, 0, ...)^{\intercal}$, which is Euclidean space $\mathbb{R}^{d}$.
Points $z\in\mathbb{R}^{d}$ are projected onto the Hyperboloid by modifying the first coordinate:
\begin{equation} \label{eq:project_up}
z_{0} \mapsto \sqrt{1+\sum_{i=1}^{d} z_{1}^{2}}
\end{equation}
and the remaining coordinates $z_{1}, ..., z_{d}$ remain the same.
The technique is computationally cheap and previously produced similar results to wrapping using an exponential transformation~\cite{macaulay2023fidelity, nagano2019wrapped}.

\section{Algorithm}\label{sec:algorithm}

We are now set up to describe our algorithm.
First, we embed genetic sequences as points (or continuous distributions for VI) in hyperbolic space using Hydra+.
Then we work with the embedded data to optimise the tree (or tree distribution).
From a set of embedded points, we compute the neighbour joining tree and compute the cost function $C$ (e.g. the phylogenetic likelihood or SIWAE) on that tree.
The overall goal is to maximise the cost function by optimising the embedding parameters (locations or variational distributions).

\subsection{Differentiable Optimisation in Tree Space}
We compute the gradient of the cost function $C$ with respect to the embedding parameters using automatic differentiation.
Automatic differentiation tracks every arithmetic operation in a numerical procedure to provide the analytical derivative of the procedure.
From the $n$ embedding locations $\vec{z}_{i} \in \mathbb{H}^{d}$ we compute the pairwise distances $D$, then the neighbour joining tree in the space of trees $T\in \mathcal{T}^{n}$, which has branch lengths that feed into the objective function $C \in \mathbb{R}$:
\begin{equation} \label{eqn:transforms}
(\mathbb{H}^{d})^{n} \xrightarrow{d_{\kappa}} \mathbb{R}^{n \choose 2} \xrightarrow{\text{soft-NJ}} \mathcal{T}^{n} \xrightarrow{C} \mathbb{R}.
\end{equation}
Automatic differentiation computes the chain rule through this series of procedures to guide the optimiser.
The impasse is that neighbour-joining is not a differentiable algorithm since it selects taxa recursively.
Below we present a differentiable version of neighbour joining based on the soft-sort algorithm.

\subsection{Soft-NJ}\label{sec:algorithm:soft}
From a set of $n$ leaf locations $\{u_{i}\}_{i=1}^{n}$ on the hyperboloid, we decode a tree using soft neighbour-joining --- passing gradients from leaf locations into branch lengths on the tree.
Neighbour-joining proceeds by recursively connecting the \textit{closest} two taxa according to the arg-min of~\cite{saitou1987neighborjoining}
\be
Q_{ij} = (n-2) d(u_{i}, u_{j}) - \sum_{k} d(u_{i}, u_{k}) - \sum_{k} d(u_{j}, u_{k}).
\ee

To select this minimum in a differentiable manner, we make use of the soft-sort algorithm \cite{prillo2020softsort}.
Soft-sort is a continuous relaxation of the arg-sort operator on a vector with a temperature parameter $\tau$ that controls the degree of approximation and impacts the gradient flow throughout the optimisation.
A colder temperature, closer to zero, reverts the soft-NJ algorithm back to the discrete (hard) version.

We use Soft-sort to create a relaxed permutation matrix of the flattened upper-triangle component $\vec{Q}$ of the $Q$ matrix as follows:
\be
P = \text{softmax}\Big(\dfrac{-|\text{sort}(\vec{Q}) \mathbbm{1}^{T} -  \mathbbm{1} \vec{Q}^{T}|}{\tau}\Big)
\ee
where $\mathbbm{1}$ is a vector of ones.
To extract the arg-min of $\vec{Q}$ we simply multiply by the last column of the permutation matrix $P$ by the vector $[1, 2, 3, ...]^{T}$.
This leads to a one-hot vector indexing the arg-min of $\vec{Q}$, which is easily unravelled into row and column one-hot vectors to use in neighbour joining.
Each of these steps is differentiable, allowing gradients to pass from $Q$ into the branch lengths on the decoded tree $T$.

In a small extension to the algorithm, we break any possible ties in $P$ by performing Soft-Sort twice.
We break ties differentiably by selecting the first minimum element of $\vec{Q}$ using the cumulative sum function.
After obtaining the permutation matrix $P$, we extract its last column denoted $P^{l}$.
We then apply soft-sort to $P^{l}C$, where $C$ is the cumulative sum $C_{i} = \sum_{k=1}^{i} P^{l}_{k}$.
This modification ensures that the first minimum element in $P^{*}$ is selected, guaranteeing a well-defined output.

\subsection{Change of Variables Jacobian}
In light of Eq.~\ref{eqn:transforms}, we are sampling trees by changing variables from $\mathbb{H}^{d\times n}$ to $\mathcal{T}^{n}$.
To account for density changes, we must include the determinant of each transformation before $\mathbb{T}^{n}$.
These changes are for sampling in $\mathbb{H}^{d\times n}$ (which is a projection from Euclidean Space as in~\cite{macaulay2023fidelity, chowdhary2018improved}), transforming by $d_{\kappa}$ (which has no associated Jacobian), and transforming by $\text{soft-NJ}$.
The Jacobian of neighbour-joining is analytically non-trivial because of the recursive nature of the algorithm.
However, the Jacobian of this series of transformations with $\text{soft-NJ}$ is easily computed using automatic differentiation.

\section{Implementation}\label{sec:implementation}
This algorithm is implemented in Dodonaphy, a software for phylogenetic inference via hyperbolic embeddings.
It uses several Python packages, notably, PyTorch for automatic differentiation~\cite{paszke2019pytorch} and DendroPy for some tree handling~\cite{sukumaran2010dendropy}.
Dodonaphy is freely available at \url{https://github.com/mattapow/dodonaphy}.
It has an easy to use command line interface and example input data for analysis.

The second release of Dodonaphy, which focuses on using gradient-based inference is available on Zenodo at: \url{https://doi.org/10.5281/zenodo.8357888}.
Additionally, the results and figures can be reproduced using the scripts available at: \url{https://github.com/mattapow/vi-fig-scripts}.

\section{Discussion}\label{sec:discus}
In this section we will demonstrate the empirical performance of gradient-based tree inference using soft-NJ.
We will evaluate its performance for both maximum likelihood and variational inference.

We have selected eight standard benchmark datasets in phylogenetics taken from~\cite{lakner2008efficiency, whidden2020systematic}.
These datasets are DNA and RNA multiple sequence alignments with between 27 and 64 tip nodes.

\subsection{Maximum Likelihood Optimisation}\label{sec:discus:maximum}
We compared the performance of our proposed hyperbolic embedding technique against two state-of-the-art maximum likelihood phylogenetic programs: IQ-TREE and RAxML-NG.

We initialise an embedding in $\mathbb{H}^{3}$ with curvature $\kappa = -100$ by embedding the BioNJ tree distances~\cite{gascuel1997bionj}.
We did this by following the hyperbolic multi-dimensional scaling approach of Hydra+~\cite{keller-ressel2020hydra}.
We then optimise the embedding locations, the curvature, and the parameters of the GTR Markov model for $2\,000$ epochs.

\begin{figure}[htbp]
\begin{center}
\includegraphics[width=.5\linewidth]{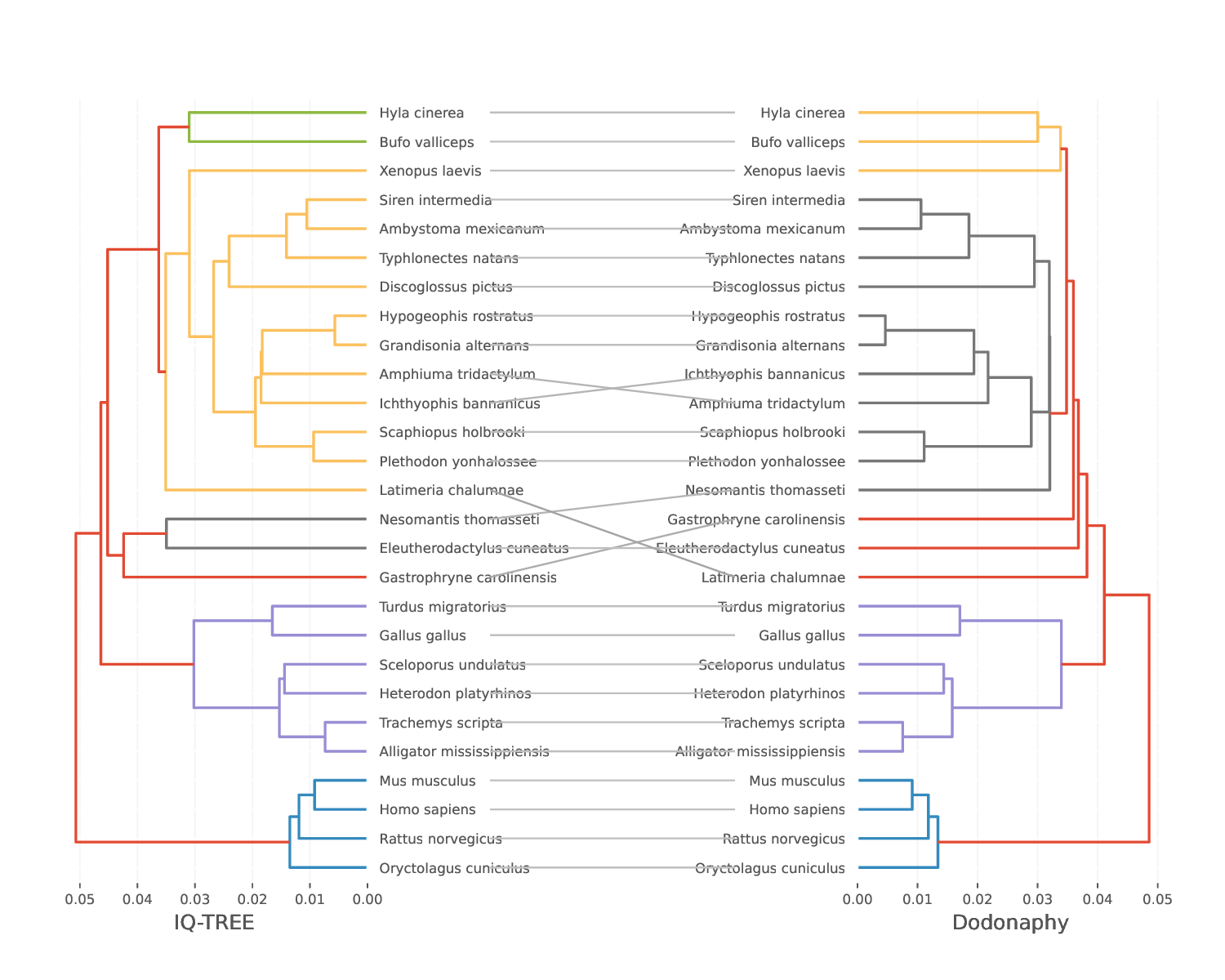}
\caption{Maximum likelihood tree found by IQ-TREE compared to Dodonaphy for data set 1.
}
\label{fig:IQ-TREE}
\end{center}
\end{figure}

Figure~\ref{fig:IQ-TREE} compares the final tree found for DS1 to IQ-TREE.
Although the resulting tree is generally similar to IQ-TREE, there are notable differences.
Both the topology and, on close inspection, branch lengths are slightly different.
It's possible that the continuous parameters are not fully optimised by Dodonaphy because it is simultaneously dealing with optimising over tree topologies in the embedding space.
To address this we propose a hybrid approach called Dodonaphy+ where we take the tree that Dodonaphy produces and optimise its continuous parameters using the BFGS optimiser available in IQ-TREE.

\begin{figure}[htbp]
\begin{center}
\includegraphics[width=.5\linewidth]{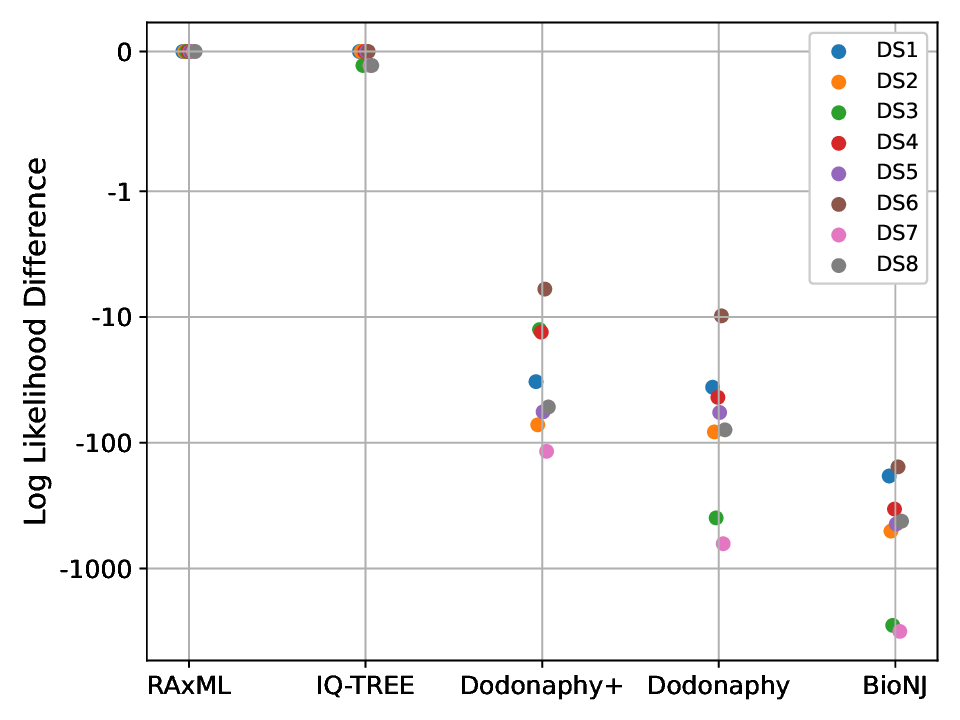}
\caption{
Difference in maximum log likelihood estimates compared to RAxML across all datasets DS1-8.
The vertical axis is negative logarithmic below $-1$ and linear above it.
}
\label{fig:ml_ds}
\end{center}
\end{figure}

To summarise these differences for all datasets we present the log likelihood under the model in table~\ref{fig:ml_ds}.
Dodonaphy consistently outperformed BioNJ demonstrating Dodonaphy's ability to improve the likelihood.
Note that the (negative) log-scale on the vertical axis downplays the significantly poorer performance of BioNJ.
Dodonaphy+ improves the maximum likelihood compared to the original Dodonaphy to varying degrees.
In DS5 the improvement is slight ($0.7$) but the change is significant for DS7 ($518.8$).

We note that after setting the curvature at $\kappa=-100$, the final curvatures across all data sets ranged from $-58.28$ (DS1) to $-75.6$ (DS3).
Previous works have quantified the tree-like of phylogenetic data~\cite{holland2002delta} as well as the relationship between curvature and the error on the four-point condition~\cite{wilson2021learning}.
These values all fall in the acceptable range previously found on these datasets~\cite{macaulay2023fidelity}.
Allowing the curvature to freely change in the optimisation process avoids imposing an arbitrary value.

\subsection{Geometric Frustration}
In practice, the state-of-the-art methods still attain better maximum likelihood estimates, indicating that the optimisation process attains a non-global optimum.
To overcome this issue, stochastic algorithms like Adam and stochastic gradient descent are commonly employed.
In this case, non-global optima can be interpreted as geometrically frustrated embedding sets, where the path to the global optimum is not along a monotone path.
However, because the tree structure is associated with the embedding, whole sets of taxa could be rearranged whilst leaving the decoded tree unchanged.
The appeal of doing this is the potentially altered neighbourhood of trees after rearrangement, providing a way out of the local optima.

One way to escape such optima would be to re-embed the tree in a new configuration in a way that preserves the outputted tree.
This is the pre-image of a given tree under neighbour joining.
For example, isometries of hyperbolic space itself are generated by the Lorentz group and (by definition) will lead to points decoding to identical trees.
However, these do not alleviate the embedding frustration.

Exploring any other embeddings in the pre-image of a tree could produce less geometrically frustrated embeddings that can then continue to be optimised.
For example, swapping the locations of two cherries could decode to the same tree.
Algebraic structures on trees~\cite{francis2022brauer} may shed some light on this, however, determining the full pre-image of neighbour joining from the embedding space is, to our knowledge, an open question.

\subsection{Variational Bayesian Inference}
Next, we use embedded distributions of trees to perform variational inference over the space of phylogenies.
We take the tip-tip distances from the IQ-TREE and embed each taxon using Hydra+.
We then associate each taxon location with a variational distribution centred at this point.
The distributions are multivariate Normals in the tangent space of the origin projected by Eq.\ref{eq:project_up}.
We optimise the parameters of these variational distributions and a point estimate of the GTR model parameters to minimise the SIWAE.
After optimising the SIWAE for $200$ epochs, we drew $10^{4}$ tree samples from the final variational distribution.

\subsubsection{Parameter Estimation}\label{sec:disuss:variational}
We compared our results to the state-of-the-art Metropolis Coupled Markov Chain Monte Carlo (MC$^{3}$) phylogenetic software MrBayes~\cite{ronquist2003mrbayes}.
We ran MrBayes with one cold chain and three heated chains for $10^{7}$ iterations.
We sampled $10^{4}$ trees evenly throughout this run as an approximation of the posterior and discarded the first $10\%$.
We use the same prior and likelihood models as in MrBayes for a fair comparison between posterior probabilities.

\begin{figure}[htbp]
\begin{center}
\includegraphics[width=.4\linewidth]{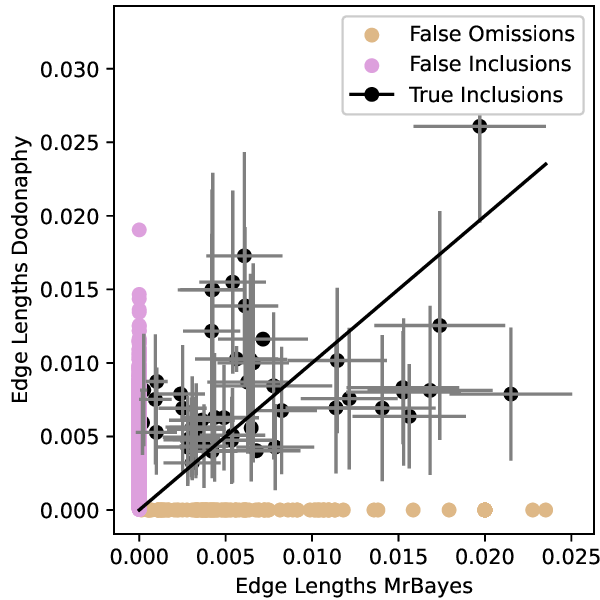}
\includegraphics[width=.53\linewidth]{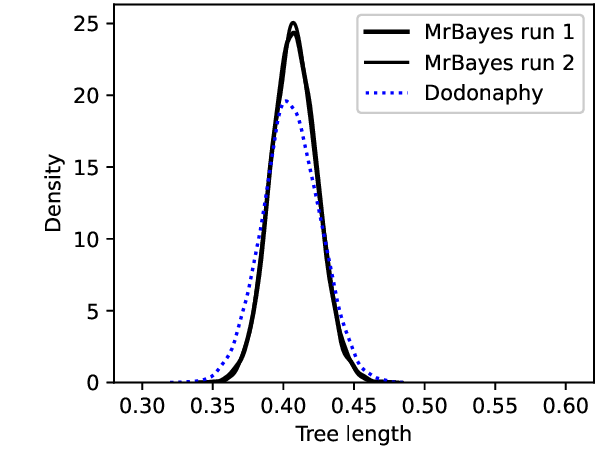}
\caption{
Variational approximation in $\mathbb{H}^{3}$ compared to MCMC.
Comparison of the split lengths (left), showing internal splits: red diamonds, and leaf splits: blue circles.
Marker opacity is set by the frequency of the split in MrBayes' estimate of the posterior.
Total tree length (kernel density) estimates (right) in the final samples.}
\label{fig:vi_estimation}
\end{center}
\end{figure}

The results show moderate agreement between the branch lengths of the posterior, figure~\ref{fig:vi_estimation}.
The estimated split frequencies and total tree lengths compare reasonably to MrBayes when considering the standard errors shown.
An exact match is not expected since VI is an approximating algorithm.
The support of the inferred tree length closely resembles that of MrBayes, although it is slightly more diffuse.

\subsubsection{Performance Evaluation}

We evaluated the performance of Dodonaphy in comparison to several state-of-art inference techniques in variational Bayesian phylogenetics.
We build on a summary of the results recently compiled in~\cite{mimori2023geophy} on the same eight datasets.
For this section we used the same model of evolution (Jukes-Cantor~\cite{jukes1969evolution}) and prior distribution used in these comparisons.
The prior is uniform across tree topologies and exponential $\text{Exp}(10)$ in the branch lengths.
We initialised Dodonaphy to the maximum likelihood tree from IQ-TREE before running the optimisation.

Then we estimated the marginal likelihood of the data over the phylogenetic parameters $\theta$ using variational Bayesian importance sampling~\cite{fourment202019}:
\be
p(D) = \int p(D|\theta) p(\theta) d\theta.
\ee
This estimator uses the variational distribution as an importance distribution for importance sampling:
\be
\hat{p}(D) = \dfrac{1}{N} \sum_{i=1}^{N} \dfrac{p(D| \tilde{\bm\theta}_{i}) p(\tilde{\bm\theta}_{i})}{q(\tilde{\bm\theta}_{i})},
\ee
where $q(\tilde{\bm\theta}_{i})$ is the variational distribution and $\tilde{\bm\theta}_{i} \sim q(\tilde{\bm\theta})$.
We used $N=1000$ samples from the variational distribution to compute this marginal estimator.

Table~\ref{tab:marginal} presents a comparison of Dodonaphy with state-of-the-art variational inference methods.
The results from stepping stone MCMC in MrBayes is also included as a baseline comparison.
Note that while VBPI-GNN has excellent results it is given topologies as inputs rather than performing topological inference.
Geophy and $\phi$-CSMC are the current state-of-art implementations performing topological and continuous parameter phylogenetic inference.

\begin{table*}
\centering
\caption{Comparison of marginal log-likelihood estimates.}
\label{tab:marginal}
\footnotesize
\begin{tabular}{lrrrrrrrr}
\hline
Dataset & DS1 & DS2 & DS3 &  DS4 & DS5 & DS6 & DS7 & DS8  \\
\hline
MrBayes & -7\,108.42 & -26\,367.57 & -33\,735.44 & -13\,330.06 & -8\,214.51 & -6\,724.07 & -37\,332.76 & -8\,649.88\\
VBPI-GNN & -7\,108.41 & -26\,367.73 & -33\,735.12 & -13\,329.94 & -8\,214.64 & -6\,724.37 & -37\,332.04 & -8\,650.65\\
Geophy LOO(3)+ & -7\,116.09 & -26\,368.54 & -33\,735.85 & -13\,337.42 & -8\,233.89 & -6\,735.90 & -37\,358.96 & -8\,660.48\\
$\phi$-CSMC & -7\,290.36 & -30\,568.49 & -33\,798.06 & -13\,582.24 & -8\,367.51 & -7\,013.83 & - & -9\,209.18 \\
Dodonaphy & -7\,006.05 & -25\,786.58 & -32\,982.86 & -12\,862.52 & -7\,211.90 & -7\,054.37 & -37\,804.35 & -9\,605.74 \\
\hline
\end{tabular}
\end{table*}

Dodonaphy generally provides poorer estimates of the posterior than competing methods.
Unlike the other phylogenetic variational techniques, the marginal log-likelihood was overestimated by Dodonaphy in some of the datasets.
This is consistent with a variational approximation that is concentrated on regions of heightened likelihood.
It's possible that after initialisation at the embedded maximum likelihood tree, the variational distribution optimised into a local optima.

The suboptimal results could also be attributed to the continuous hyperbolic variational approximation.
Underlying this model is the assumption that trees with similar tip-tip distances share similar posterior likelihoods.
This assumption is a heuristic that provides an efficient way to encode tree distributions but may constrain the flexibility of the distribution.
These findings are also consistent with a variational distribution that is too simple, calling for a more expressiveness.
We explore this by boosting the variational distribution.

\subsection{Effect of Boosting}\label{sec:discuss:boosting}
\begin{figure}[htbp]
\begin{center}
\includegraphics[width=.5\linewidth]{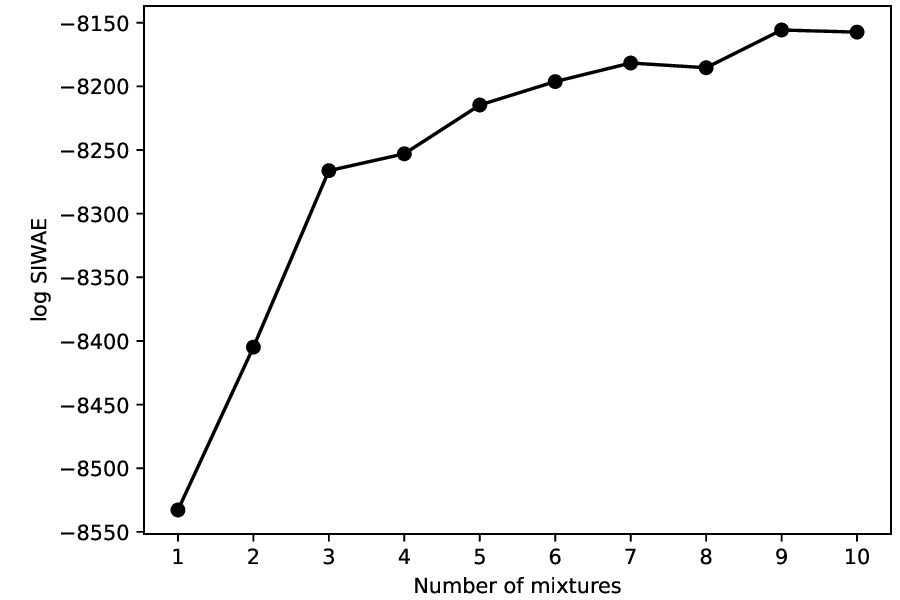}
\caption{
Effect of the number of boosts on the final SIWAE estimate 
for DS1.
}
\label{fig:boosts}
\end{center}
\end{figure}

Whilst boosting improves the expressiveness of the variational distribution, it also increases the computational demand of variational inference by a factor of $K$, so we are interested in the minimal number of mixtures required.
To understand the number of boosts required to capture the embedded posterior distribution of trees, we fixed the number of importance samples at $M=3$ and varied the number of mixtures $K$ from one to ten.
We optimised for $200$ epochs starting from the IQ-TREE distances.
The final SIWAE value suggests that the presence of additional mixtures improves the variational approximation, although the improvement slowly saturates after $M = 3$, figure~\ref{fig:boosts}.
Having this flexible variational family increases the inference accuracy and opens up more complex tree distributions.

\subsection{Outlook}
Hyperbolic tree embeddings, through the use of soft-NJ, provide a differentiable way to efficiently encode trees and even distributions of trees.
This advancement paves the way for continuous optimisation over low-dimensional representations of tree spaces.
It opens up differentiable methods for a broad range of inference techniques to tackle phylogenetics.

We demonstrated two applications in maximum likelihood and variational inference.
However, the challenges of non-convexity and poor variational approximations pose open challenges to the research community to fully realise the potential of hyperbolic tree optimisation.
Notably, finding the pre-image of the decoding process could alleviate geometric frustration aiding optimisation challenges.
Additionally, exploring alternative approximating functions or transitioning to full-rank variational approximation may increase the variational quality of this approach.

\section{Competing interests}
No competing interest is declared.

\section{Author contributions statement}

M.M. and M.F conceived and analysed the experiments.
M.M. conducted the experiments and wrote the manuscript.
M.F. reviewed the manuscript.

\section{Acknowledgments}
The authors thank the reviewers for their valuable suggestions.

This work was supported by the Australian Government through the Australian Research Council (project number LP180100593).

Computational facilities were provided by the UTS eResearch High Performance Computer Cluster.

\bibliographystyle{plain}
\bibliography{main}

\begin{thebibliography}{10}

\bibitem{allman2008phylogenetic}
Elizabeth~S. Allman and John~A. Rhodes.
\newblock Phylogenetic ideals and varieties for the general {{Markov}} model.
\newblock {\em Advances in Applied Mathematics}, 40(2):127--148, February 2008.

\bibitem{billera2001geometry}
Louis~J. Billera, Susan~P. Holmes, and Karen Vogtmann.
\newblock Geometry of the {{Space}} of {{Phylogenetic Trees}}.
\newblock {\em Advances in Applied Mathematics}, 27(4):733--767, November 2001.

\bibitem{blei2017variational}
David~M. Blei, Alp Kucukelbir, and Jon~D. McAuliffe.
\newblock Variational {{Inference}}: {{A Review}} for {{Statisticians}}.
\newblock {\em Journal of the American Statistical Association},
  112(518):859--877, April 2017.

\bibitem{burda2016importance}
Yuri Burda, Roger Grosse, and Ruslan Salakhutdinov.
\newblock Importance {{Weighted Autoencoders}}, November 2016.

\bibitem{chami2020trees}
Ines Chami, Albert Gu, Vaggos Chatziafratis, and Christopher R{\'e}.
\newblock From trees to continuous embeddings and back: {{Hyperbolic}}
  hierarchical clustering.
\newblock In H.~Larochelle, M.~Ranzato, R.~Hadsell, M.~F. Balcan, and H.~Lin,
  editors, {\em Advances in Neural Information Processing Systems}, volume~33,
  pages 15065--15076. {Curran Associates, Inc.}, 2020.

\bibitem{chami2020lowdimensional}
Ines Chami, Adva Wolf, Frederic Sala, and Christopher R{\'e}.
\newblock Low-{{Dimensional Knowledge Graph Embeddings}} via {{Hyperbolic
  Rotations}}.
\newblock In {\em {{NeurIPS}}}, volume~10, page~v1, {Vancouver, Canada}, 2020.

\bibitem{chowdhary2018improved}
Kenny Chowdhary and Tamara~G Kolda.
\newblock An improved hyperbolic embedding algorithm.
\newblock {\em Journal of Complex Networks}, 6(3):321--341, July 2018.

\bibitem{corso2021neural}
Gabriele Corso, Zhitao Ying, Michal P{\'a}ndy, Petar Veli{\v c}kovi{\'c}, Jure
  Leskovec, and Pietro Li{\`o}.
\newblock Neural {{Distance Embeddings}} for {{Biological Sequences}}.
\newblock In {\em Advances in {{Neural Information Processing Systems}}},
  volume~34, pages 18539--18551. {Curran Associates, Inc.}, 2021.

\bibitem{dinh2017probabilistic}
Vu~Dinh, Arman Bilge, Cheng Zhang, and Frederick~A Matsen.
\newblock Probabilistic {{Path Hamiltonian Monte Carlo}}.
\newblock In {\em Machine {{Learning}}}, volume~70 of {\em Proceedings of
  {{Machine Learning Research}}}, page~10, {International Convention Centre,
  Sydney, Australia}, 2017. {PMLR}.

\bibitem{felsenstein1973maximum}
Joseph Felsenstein.
\newblock Maximum {{Likelihood}} and {{Minimum-Steps Methods}} for {{Estimating
  Evolutionary Trees}} from {{Data}} on {{Discrete Characters}}.
\newblock {\em Systematic Biology}, 22(3):240--249, September 1973.

\bibitem{fourment202019}
Mathieu Fourment, Andrew~F Magee, Chris Whidden, Arman Bilge, Frederick~A
  Matsen~IV, and Vladimir~N Minin.
\newblock 19 dubious ways to compute the marginal likelihood of a phylogenetic
  tree topology.
\newblock {\em Systematic biology}, 69(2):209--220, 2020.

\bibitem{francis2022brauer}
Andrew Francis and Peter~D. Jarvis.
\newblock Brauer and partition diagram models for phylogenetic trees and
  forests.
\newblock {\em Proceedings of the Royal Society A: Mathematical, Physical and
  Engineering Sciences}, 478(2262):20220044, June 2022.

\bibitem{gascuel1997bionj}
O~Gascuel.
\newblock {{BIONJ}}: An improved version of the {{NJ}} algorithm based on a
  simple model of sequence data.
\newblock {\em Molecular Biology and Evolution}, 14(7):685--695, July 1997.

\bibitem{guindon2010new}
St{\'e}phane Guindon, Jean-Fran{\c c}ois Dufayard, Vincent Lefort, Maria
  Anisimova, Wim Hordijk, and Olivier Gascuel.
\newblock New {{Algorithms}} and {{Methods}} to {{Estimate Maximum-Likelihood
  Phylogenies}}: {{Assessing}} the {{Performance}} of {{PhyML}} 3.0.
\newblock {\em Systematic Biology}, 59(3):307--321, May 2010.

\bibitem{holland2002delta}
B.~R. Holland, K.~T. Huber, A.~Dress, and V.~Moulton.
\newblock Delta plots: A tool for analyzing phylogenetic distance data.
\newblock {\em Molecular Biology and Evolution}, 19(12):2051--2059, December
  2002.

\bibitem{iuchi2021representation}
Hitoshi Iuchi, Taro Matsutani, Keisuke Yamada, Natsuki Iwano, Shunsuke Sumi,
  Shion Hosoda, Shitao Zhao, Tsukasa Fukunaga, and Michiaki Hamada.
\newblock Representation learning applications in biological sequence analysis.
\newblock {\em bioRxiv}, page 2021.02.26.433129, February 2021.

\bibitem{jukes1969evolution}
Thomas~H Jukes and Charles~R Cantor.
\newblock Evolution of protein molecules.
\newblock {\em Mammalian protein metabolism}, 3:21--132, 1969.

\bibitem{keller-ressel2020hydra}
Martin {Keller-Ressel} and Stephanie Nargang.
\newblock Hydra: A method for strain-minimizing hyperbolic embedding of
  network- and distance-based data.
\newblock {\em Journal of Complex Networks}, 8(1):cnaa002, February 2020.

\bibitem{ki2022variational}
Caleb Ki.
\newblock Variational {{Phylodynamic Inference Using Pandemic-scale Data}}.
\newblock {\em Molecular Biology and Evolution}, 39, August 2022.

\bibitem{koptagel2022vaiphy}
Hazal Koptagel, Oskar Kviman, Harald Melin, Negar Safinianaini, and Jens
  Lagergren.
\newblock {{VaiPhy}}: A variational inference based algorithm for phylogeny.
\newblock In S.~Koyejo, S.~Mohamed, A.~Agarwal, D.~Belgrave, K.~Cho, and A.~Oh,
  editors, {\em Advances in Neural Information Processing Systems}, volume~35,
  pages 14758--14770. {Curran Associates, Inc.}, 2022.

\bibitem{lakner2008efficiency}
Clemens Lakner, Paul {van der Mark}, John~P. Huelsenbeck, Bret Larget, and
  Fredrik Ronquist.
\newblock Efficiency of {{Markov Chain Monte Carlo Tree Proposals}} in
  {{Bayesian Phylogenetics}}.
\newblock {\em Systematic Biology}, 57(1):86--103, February 2008.

\bibitem{layer2017phylogenetic}
Mark Layer and John~A. Rhodes.
\newblock Phylogenetic trees and {{Euclidean}} embeddings.
\newblock {\em Journal of Mathematical Biology}, 74(1-2):99--111, January 2017.

\bibitem{macaulay2023fidelity}
Matthew Macaulay, Aaron Darling, and Mathieu Fourment.
\newblock Fidelity of hyperbolic space for {{Bayesian}} phylogenetic inference.
\newblock {\em PLOS Computational Biology}, 19(4):e1011084, April 2023.

\bibitem{mimori2023geophy}
Takahiro Mimori and Michiaki Hamada.
\newblock {{GeoPhy}}: {{Differentiable Phylogenetic Inference}} via {{Geometric
  Gradients}} of {{Tree Topologies}}, July 2023.

\bibitem{minh2020iqtree}
Bui~Quang Minh, Heiko~A Schmidt, Olga Chernomor, Dominik Schrempf, Michael~D
  Woodhams, Arndt {von Haeseler}, and Robert Lanfear.
\newblock {{IQ-TREE}} 2: {{New Models}} and {{Efficient Methods}} for
  {{Phylogenetic Inference}} in the {{Genomic Era}}.
\newblock {\em Molecular Biology and Evolution}, 37(5):1530--1534, May 2020.

\bibitem{monath2019gradientbased}
Nicholas Monath, Manzil Zaheer, Daniel Silva, Andrew McCallum, and Amr Ahmed.
\newblock Gradient-based {{Hierarchical Clustering}} using {{Continuous
  Representations}} of {{Trees}} in {{Hyperbolic Space}}.
\newblock In {\em Proceedings of the 25th {{ACM SIGKDD International
  Conference}} on {{Knowledge Discovery}} \& {{Data Mining}}}, pages 714--722,
  {Anchorage AK USA}, July 2019. {ACM}.

\bibitem{morningstar2021automatic}
Warren Morningstar, Sharad Vikram, Cusuh Ham, Andrew Gallagher, and Joshua
  Dillon.
\newblock Automatic {{Differentiation Variational Inference}} with
  {{Mixtures}}.
\newblock In {\em Proceedings of {{The}} 24th {{International Conference}} on
  {{Artificial Intelligence}} and {{Statistics}}}, pages 3250--3258. {PMLR},
  March 2021.

\bibitem{nagano2019wrapped}
Yoshihiro Nagano, Shoichiro Yamaguchi, Yasuhiro Fujita, and Masanori Koyama.
\newblock A {{Wrapped Normal Distribution}} on {{Hyperbolic Space}} for
  {{Gradient-Based Learning}}.
\newblock In {\em International {{Conference}} on {{Machine Learning}}}, pages
  4693--4702. {PMLR}, May 2019.

\bibitem{nickel2017poincare}
Maximillian Nickel and Douwe Kiela.
\newblock Poincar\'e {{Embeddings}} for {{Learning Hierarchical
  Representations}}.
\newblock In {\em Advances in {{Neural Information Processing Systems}}},
  volume~30, pages 6338--6347, 2017.

\bibitem{paszke2019pytorch}
Adam Paszke, Sam Gross, Francisco Massa, Adam Lerer, James Bradbury, Gregory
  Chanan, Trevor Killeen, Zeming Lin, Natalia Gimelshein, Luca Antiga, et~al.
\newblock Pytorch: An imperative style, high-performance deep learning library.
\newblock {\em Advances in neural information processing systems}, 32, 2019.

\bibitem{peng2022hyperbolic}
Wei Peng, Tuomas Varanka, Abdelrahman Mostafa, Henglin Shi, and Guoying Zhao.
\newblock Hyperbolic {{Deep Neural Networks}}: {{A Survey}}.
\newblock {\em IEEE Transactions on Pattern Analysis and Machine Intelligence},
  44(12):10023--10044, December 2022.

\bibitem{prillo2020softsort}
Sebastian Prillo and Julian Eisenschlos.
\newblock {{SoftSort}}: {{A Continuous Relaxation}} for the argsort
  {{Operator}}.
\newblock In {\em Proceedings of the 37th {{International Conference}} on
  {{Machine Learning}}}, pages 7793--7802. {PMLR}, November 2020.

\bibitem{rannala2012tail}
Bruce Rannala, Tianqi Zhu, and Ziheng Yang.
\newblock Tail {{Paradox}}, {{Partial Identifiability}}, and {{Influential
  Priors}} in {{Bayesian Branch Length Inference}}.
\newblock {\em Molecular Biology and Evolution}, 29(1):325--335, January 2012.

\bibitem{ronquist2003mrbayes}
F.~Ronquist and J.~P. Huelsenbeck.
\newblock {{MrBayes}} 3: {{Bayesian}} phylogenetic inference under mixed
  models.
\newblock {\em Bioinformatics}, 19(12):1572--1574, August 2003.

\bibitem{saitou1987neighborjoining}
N~Saitou and M~Nei.
\newblock The neighbor-joining method: A new method for reconstructing
  phylogenetic trees.
\newblock {\em Molecular Biology and Evolution}, 4(4):406--425, July 1987.

\bibitem{speyer2003tropical}
David Speyer and Bernd Sturmfels.
\newblock The {{Tropical Grassmannian}}.
\newblock {\em Advances in Geometry}, 4:389--–411, 2004.

\bibitem{stamatakis2014raxml}
Alexandros Stamatakis.
\newblock {{RAxML}} version 8: A tool for phylogenetic analysis and
  post-analysis of large phylogenies.
\newblock {\em Bioinformatics}, 30(9):1312--1313, May 2014.

\bibitem{sukumaran2010dendropy}
Jeet Sukumaran and Mark~T. Holder.
\newblock {{DendroPy}}: A {{Python}} library for phylogenetic computing.
\newblock {\em Bioinformatics}, 26(12):1569--1571, June 2010.

\bibitem{sumner2017dimensional}
Jeremy~G. Sumner.
\newblock Dimensional {{Reduction}} for the {{General Markov Model}} on
  {{Phylogenetic Trees}}.
\newblock {\em Bulletin of Mathematical Biology; New York}, 79(3):619--634,
  March 2017.

\bibitem{tavare1986probabilistic}
Simon Tavare.
\newblock Some probabilistic and statistical problems in the analysis of
  {{DNA}} sequences.
\newblock {\em American Mathematical Society}, 17(2):57--86, 1986.

\bibitem{tran2021variational}
Minh-Ngoc Tran, Dang~H. Nguyen, and Duy Nguyen.
\newblock Variational {{Bayes}} on manifolds.
\newblock {\em Statistics and Computing}, 31(6):71, September 2021.

\bibitem{whidden2020systematic}
Chris Whidden, Brian~C Claywell, Thayer Fisher, Andrew~F Magee, Mathieu
  Fourment, and Frederick~A Matsen, IV.
\newblock Systematic {{Exploration}} of the {{High Likelihood Set}} of
  {{Phylogenetic Tree Topologies}}.
\newblock {\em Systematic Biology}, 69(2):280--293, March 2020.

\bibitem{wilson2021learning}
Benjamin Wilson.
\newblock Learning phylogenetic trees as hyperbolic point configurations.
\newblock {\em arXiv:2104.11430 [cs]}, April 2021.

\bibitem{wilson2018gradient}
Benjamin Wilson and Matthias Leimeister.
\newblock Gradient descent in hyperbolic space.
\newblock {\em arXiv:1805.08207 [math]}, August 2018.

\bibitem{zhang2020improved}
Cheng Zhang.
\newblock Improved {{Variational Bayesian Phylogenetic Inference}} with
  {{Normalizing Flows}}.
\newblock In D.~Anderson, editor, {\em Neural {{Information Processing
  Systems}}}, pages 22--30, {Vancouver, Canada}, 2020. {American Institute of
  Physics}.

\bibitem{zhang2019variational}
Cheng Zhang and Frederick~A Matsen.
\newblock Variational {{Bayesian}} phylogenetic inference.
\newblock In {\em International {{Conference}} on {{Learning
  Representations}}}, page~15, 2019.

\end{thebibliography}

\end{document}